\documentclass[aps,twocolumn,showpacs]{revtex4}
\usepackage{graphicx,ulem}

\begin{document}

\title{Effect of optical disorder and single defects on the expansion of
a Bose-Einstein condensate in a one-dimensional waveguide}

\author{C.~Fort, L.~Fallani, V.~Guarrera, J.~Lye, M.~Modugno$^1$, D.~S.~Wiersma$^2$, M.~Inguscio}

\affiliation{LENS,~Dipartimento~di~Fisica,~and~INFM~Universit\`a~di~Firenze,\\~via~Nello~Carrara~1,~I-50019~Sesto~Fiorentino~(FI),~Italy
\\ $^1$~also~BEC-INFM~Center,~Universit\`a~di~Trento,~I-38050~Povo~(TN),~Italy
\\  $^2$~also~INFM-MATIS,~Catania,~Italy }

\begin{abstract}
We investigate the one-dimensional expansion of a Bose-Einstein
condensate in an optical guide in the presence of a random
potential created with optical speckles. With the speckle the
expansion of the condensate is strongly inhibited. A detailed
investigation has been carried out varying the experimental
conditions and checking the expansion when a single optical defect
is present. The experimental results are in good agreement with
numerical calculations based on the Gross-Pitaevskii equation.
\end{abstract}

\pacs{03.75.Kk, 42.25.Dd, 32.80.Pj}


\maketitle

The study of Bose-Einstein condensates (BECs) in random potentials
has gained much interest in the last years. Interesting phenomena
can be studied in the context of wave transport in random systems,
such as Anderson localization in which transport is dramatically
suppressed due to disorder. Anderson localization was first
proposed to explain the metal-insulator transition in electron
transport in disordered solids \cite{anderson,ando} and later
predicted \cite{w2} and observed \cite{w3} also for other wave
phenomena such as light and sound. The fundamental concept of
Anderson localization applies to any wave phenomenon and should be
observable for ultra-cold atoms as matter waves propagating in a
random potential.

One-dimensional (1D) random systems are excellent for the
observation of Anderson localization as has been shown for optical
waves \cite{w4}. In cold Bose gases, due to the presence of
interactions and thanks to the demonstrated possibility to create
the Mott insulator state in a deep optical lattice \cite{mott},
the range of interesting and new effects increases, including the
Bose-glass phase and spin glasses \cite{parisi}. Different
theoretical papers have already addressed these problems
\cite{rothburnett,damski,miniatura,castin} and a first
experimental study on static and dynamic properties of a BEC in a
random potential created by light has been reported in
\cite{nostrorandom}.

In this paper we present an experimental study of a Bose-Einstein
condensate expanding in a 1D optical guide in the presence of a
random potential created by optical speckles. We find that in the
speckle potential the 1D expansion is strongly suppressed. In
order to understand the role played by the disorder we also study
the BEC expansion when only a single defect is present. Numerical
calculations based on the Gross-Pitaevskii equation (GPE) are in
perfect agreement with the experimental results.

We first produce a Bose-Einstein condensate of $^{87}$Rb atoms in
a Ioffe-Pritchard magnetic trap with axial and radial frequencies
$\omega_z=2\pi \times (8.74\pm0.03)$~Hz and $\omega_\perp=2\pi
\times (85\pm1)$~Hz respectively, with the axis of the trap
oriented horizontally. Our typical BECs are made of $\simeq 2
\times 10^5$ atoms in the hyperfine ground state $|F=1;m_F=-1>$.
Then we adiabatically transfer the condensates into a crossed
optical dipole trap with the same elongated symmetry as the
magnetic trap. The optical trap is realized aligning two
orthogonal beams derived from a Ti:Sa laser working at 830~nm (far
red-detuned with respect to the atomic transition at 795~nm). The
horizontal beam, shone along the axial direction of the
condensate, has a beam waist of $40$~$\mu$m and a typical power of
40~mW while the vertical beam is characterized by a beam waist of
$130$~$\mu$m and a typical power of 50~mW. The measured trapping
frequencies of the optical trap are $\omega_{OTz}=2\pi \times
(24.7 \pm 0.8)$~Hz in the axial direction and
$\omega_{OT\perp}=2\pi \times (293 \pm 6)$~Hz in the radial
direction. The two beams are derived from the same laser and pass
through Acousto Optic Modulators (AOMs) working at different
frequencies (with a difference of 7~MHz). We switch on the crossed
dipole trap adiabatically using a 200~ms exponential ramp with a
time constant of 50~ms. After an additional time interval of
100~ms we switch off the magnetic trap in order to leave the
condensate in the pure optical trap and wait 1~s to let the system
equilibrate to the ground state. The transfer efficiency from the
pure magnetic trap to the pure optical trap is $\sim$50\%,
corresponding to a condensate of $\sim$$ 10^5$ atoms with a
chemical potential $\mu/h \simeq 2.5$~kHz and typical radii
(calculated in the Thomas-Fermi regime) of $R_z=30$~$\mu$m and
$R_\perp=2.6$~$\mu$m in the axial and radial direction.

In order to induce a 1D expansion of the condensate, we switch off
abruptly (in less than 1~ms) the vertical trapping laser beam. The
horizontal laser beam results in an optical guide, the axial
frequency being  $\simeq 2\pi \times 1$~Hz. Furthermore, the
optical guide is tilted by $8.6 \pm 1.6$~mrad with respect to the
horizontal plane resulting in a measured acceleration along the
guide, $a=0.096\pm0.004$~ms$^{-2}$, caused by gravity.

We observe the expansion of the BEC in the random potential adding
optical speckles \cite{altrirandom}. The speckles are obtained and
characterized as described in our previous work
\cite{nostrorandom}. The beam  for the speckles is derived from
the same Ti:Sa laser we use to create the dipole trap, but passed
through a different AOM (with a detuning of 10 and 17~MHz from the
AOMs controlling the crossed dipole trap) in order to have an
independent control on its time switching. The speckle beam, after
passing through a diffusive plate, propagates along the horizontal
radial direction of the condensate and is collinear with the
resonant laser beam used for the imaging of the condensate. The
imaging setup can be used to detect both the BEC and the speckle
pattern, enabling us to characterize the actual random potential
experienced by the condensate \cite{nostrorandom}. The Fourier
transform of the speckle potential indicates that the smallest
length scale of the speckle is 10~$\mu$m \cite{autocorrelazione},
much bigger than the radial size of the condensate. Therefore, the
condensate experiences the varying random potential only along the
optical guide axis during expansion. We define the speckle height
$V_S$ by taking twice the standard deviation of the speckle
potential along the BEC axial direction. The speckle height and
the other energy scales will be conveniently expressed throughout
this Letter in units of frequency (using the implicit assumption
of a division by the Planck constant $h$).

\begin{figure}[h!]
\begin{center}
\includegraphics[width=0.9\columnwidth]{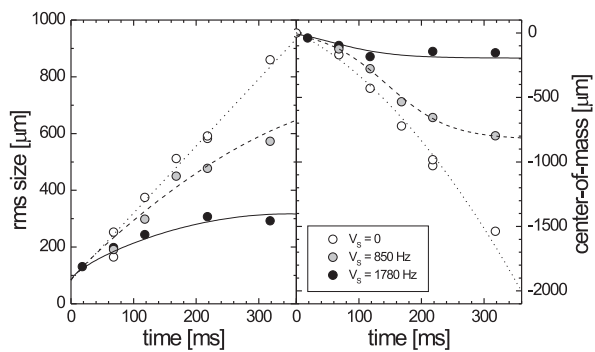}
\end{center}
\caption{A) Rms radius of the condensate expanding in the optical
linear guide + speckle potential as a function of time for three
different values of the speckle height with $V_S/\mu<1$. B)
Corresponding center of mass motion. The lines are guides for the
eyes.} \label{raggi+centri}
\end{figure}

In a first series of measurements we adiabatically switch on the
speckle potential together with the crossed dipole trap in order
to let the condensate equilibrate to the ground state of the
combined potential. In Fig.\ref{raggi+centri} we report the rms
radius and the center of mass of the atomic cloud expanding in the
optical guide once the vertical trap beam has been switched off as
a function of time for different speckle potential heights.
Without speckles the condensates freely expands, while in the
presence of the speckles both the expansion and the center of mass
motion are inhibited after few hundreds of ms for $V_S/ \mu >
0.3$.

\begin{figure}[h!]
\begin{center}
\includegraphics[width=0.8\columnwidth]{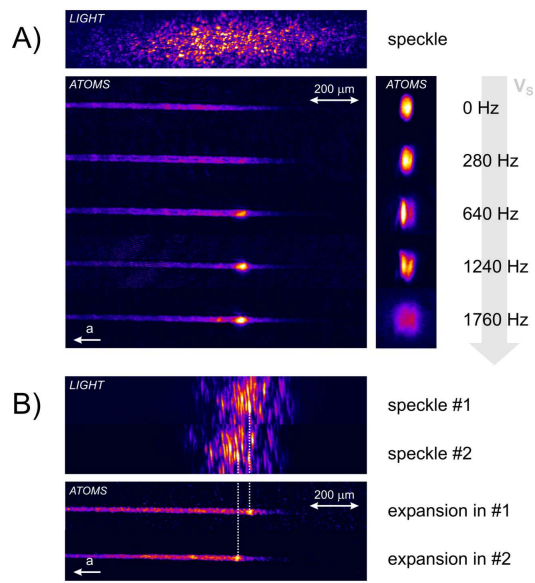}
\end{center}
\caption{A) left - in situ picture of the speckle potential and
density profiles of the condensate after 118~ms of expansion in
the linear guide for $0<V_S< 1.7$~kHz. right - density profile of
the condensate released from the crossed dipole trap + speckle
potential after 18~ms of free expansion. B) For two different
positions of the speckle field we show the density distribution of
the atomic cloud after 300~ms of expansion in the linear optical
guide with the speckle switched on abruptly and $V_S= 2$~kHz.}
\label{espansione}
\end{figure}

In Fig.\ref{espansione}A we report the density profile of the
condensate, imaged in situ in the optical guide, after 118~ms of
expansion for different speckle potential heights (ranging from
$V_S/\mu=0$ to $0.7$) together with the picture of the actual
speckle field used. A closer look shows that actually two
different components can be distinguished: while a low density
cloud expands without stopping, a few localized density peaks are
observable. In the same Figure we also show the measured density
distribution obtained releasing the BEC ground state from the
crossed dipole trap + random potential. The expansion from the
highest speckle potential of $1.8$~kHz shows a broad Gaussian
profile that is compatible with interference from separate,
randomly distributed condensates \cite{nostrorandom}, as one would
expect in the tight binding regime. For lower speckle potential
only density modulations of the Thomas-Fermi profile are observed,
indicating that for $V_S< 1.2$~kHz we are not in the tight binding
regime. However, we still observe a halting of the expansion in
the linear guide.

In order to have a further insight into the mechanism causing the
suppression of transport, we repeat the experiment either putting
the speckle pattern to one side of the crossed dipole trap center,
or switching on the speckle abruptly to reduce trapping in the
deepest speckle wells. In this latter case, we switch on the
speckle only after 50~ms of free expansion in the linear guide.
This preliminary expansion allows a reduction of the interaction
energy to 30\% of the initial value and produces a bigger axial
size corresponding to a larger number of speckle peaks ($\simeq
50$) across the condensate. Due to the acceleration along the
optical guide, when the speckle potential is added, the kinetic
energy of the condensate center of mass is $2.5$~kHz. In
Fig.\ref{espansione}B we show the density distribution of the
condensate after 300~ms of expansion for two different positions
of the same speckle realization with $V_S= 2$~kHz. We still
observe a transport inhibition that is characterized by peaks in
the density distribution that spatially shift following the
speckles potential (as shown by dashed lines in the figure). We
further investigate the correlation between the position of the
speckle potential and the localized peaks in order to understand
the role of disorder in the observed behaviour.

\begin{figure}[h!]
\begin{center}
\includegraphics[width=0.9\columnwidth]{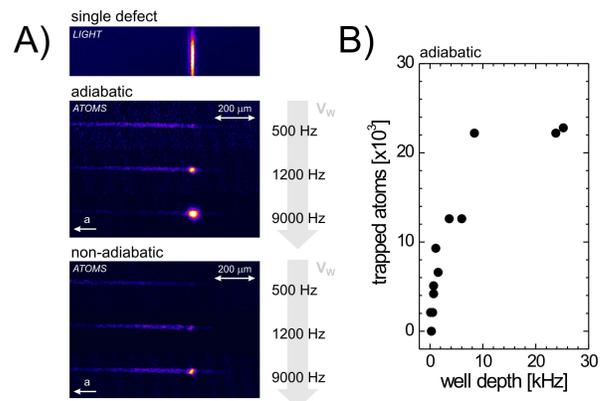}
\end{center}
\caption{A) Intensity profile of the laser beam creating the
single defect and density profiles of the condensate after 300~ms
of expansion for $ 500 $~Hz~$<V_w<  9.0$~kHz either adiabatically
or abruptly adding the defect. B) Number of atoms trapped in the
single defect as a function of $V_w$ when the potential well is
switched on adiabatically.} \label{singledefect}
\end{figure}

For this purpose, we perform a series of experiments studying the
expansion of the condensate in the linear guide when a single
defect, instead of a randomly distributed series, is present. The
single defect is obtained by removing the diffusive plate along
the beam that creates the speckles potential. We then change the
optics in order to focus the beam onto the condensate. This
creates a single optical well (with an elongated shape obtained
using a cylindrical lens) which size and depth $V_w$ is defined
through a Gaussian fit. The size of the well is $\sigma_z=6
$~$\mu$m along the horizontal direction and $\sigma_y=85$~$\mu$m
along the vertical direction. We study the expansion of the
condensate in the linear guide with the single defect with the
same time sequence used for the speckles potential: either
switching on the single defect adiabatically together with the
crossed trap, or abruptly after a preliminary expansion in the
linear guide. The observed density profiles are reported in
Fig.\ref{singledefect}A. While a low density component expands
freely, a sharp peak in the density profile can be observed
corresponding to part of the condensate trapped in the single
defect and seems very similar to the effect seen with the
speckles. The population of atoms localized in the single well
increases as a function of the well depth as shown in
Fig.\ref{singledefect}B. The efficiency of trapping is reduced
when the well is switched on abruptly but we still have some
trapping for $V_w\geq 1$~kHz, corresponding to a depth of the
single well comparable with the speckle potential that stops the
expansion of the condensate. These results indicate that in our
experiment the effect of trapping in the deepest wells of the
random potential cannot be avoided even in a regime of $V_w<\mu$.

\begin{figure}[h!]
\begin{center}
\includegraphics[width=0.8\columnwidth,clip=]{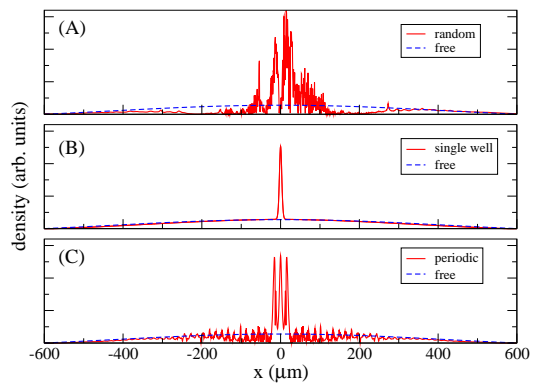}
\end{center}
\caption{Calculated density profiles after 75~ms of expansion for
random potential (A), single Gaussian well (B) and periodic
potential (C), compared with the free expansion case (dashed
line). In all three cases the potential height is 0.4~$\mu$.}
\label{teo}
\end{figure}

Our observations are supported by numerical calculations based on
the GPE. As in the experiment we consider both the case of a
random potential with correlation length $\sigma_z=$5~$\mu$m and a
single Gaussian well. Moreover, in order to better enlighten the
actual role played by randomness, we also consider the case of a
periodic potential with the same spacing and height of the
speckles. The condensate is initially prepared in the ground state
of the combined potential, and then let expand through the optical
guide (neglecting for simplicity gravity). The density profiles
after 75~ms of expansion are shown in Fig.\ref{teo}. In all the
cases two components are clearly identified: the lateral wings
that expand almost freely, and a central part that is localized in
the deepest wells of the potential. The comparison between these
different situations indicates that the  observed effect is mainly
due to deep wells in the potential acting as single traps when the
local chemical potential becomes of the order of their height.
This could mask the possible observation of other localization
effects due to the cumulative behaviour of the disordered
potential wells. The theoretical predictions are in perfect
agreement with the experimental observations. Note also that in
this regime interactions play mainly against localization since
they provide the initial energy that allows the wings to expand,
whereas the dephasing that it is produced in the central part is
only a secondary effect \cite{teoria}.

The system we have realized can be used to deeply investigate the
interaction between a condensate and a single scatterer. Anderson
localization is due to interference between partially reflected
matter-waves from randomly distributed scatterers. Although
interactions in a condensate may destroy Anderson localization,
experimental studies with a single defect can be very useful to
understand if the random potential created by light is a proper
tool to observe this effect even in absence of interactions, for
instance working with fermions.

\begin{figure}[h!]
\begin{center}
\includegraphics[width=0.8\columnwidth]{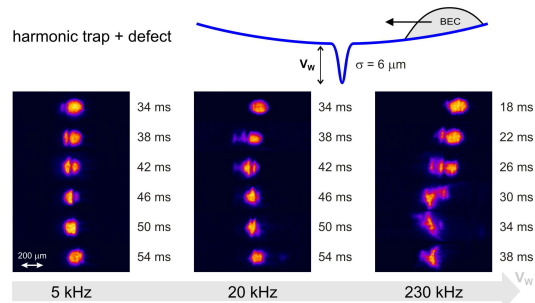}
\end{center}
\caption{top) Schematic of the experiment. bottom) Density
profiles of the condensate performing dipole oscillations in the
magnetic trap and interacting with a single well.}
\label{singleintrap}
\end{figure}

In this context we investigate the transmission-reflection of a
moving condensate ``colliding'' with a single optical defect. We
induce axial dipole oscillation of the condensate confined in the
harmonic magnetic trap \cite{nota2} adding a single optical well
as shown in Fig.\ref{singleintrap}. Varying the distance of the
defect from the center of the magnetic trap and the amplitude of
the oscillation, we can finely tune the center of mass kinetic
energy of the condensate. Considering a single particle moving at
constant velocity, in order to observe quantum reflection of the
matter wave from a single well two different conditions have to be
fulfilled \cite{ketterleQR}. The first condition regards the
comparison between the kinetic energy of the particle $E_k$ that
should be smaller than the depth of the well $|V_w|$. This
condition is the only requirement in the case of a square well,
but for different shapes of the well a second condition becomes
important. The potential should vary more than $E_k$ in a distance
short compared to the de~Broglie wavelength of the particle
$\lambda_{dB}$, i.e. the following relation should hold:
$|dV_w/dz|\; \lambda_{dB}>E_k$. When near-infrared light is used
to create the defects, this condition becomes very difficult to
fulfill. The optical access in standard BEC apparatus limits the
size of light defects to several microns resulting in the
requirement of intense light or very low condensate velocities.
Very deep optical potentials can induce heating of the atomic
cloud. In our experiment, to check the temperature of the system,
we have performed the imaging of the condensate after a free
expansion of 18~ms.

In Fig.\ref{singleintrap} we show a series of images corresponding
to the condensate performing dipole center of mass oscillations in
the harmonic magnetic potential and interacting with a single
well. In the experiment we vary the time after the excitation of
the dipole motion and the depth of the optical well. For $V_w<
200$~kHz when the condensate approaches (with a center of mass
kinetic energy $E_k \simeq 5 $~kHz) the potential well at $t\simeq
40$~ms a hole in the density distribution forms due to the fast
acceleration induced by the potential walls. The condensate
recovers its starting density distribution after crossing the
well. In this regime of parameters, as expected, we do not observe
a quantum reflected component. As we increase the depth of the
well to $V_w\gtrsim  200$~kHz the condensate starts to be
destroyed by the interaction with light. These observations
demonstrate that it seems quite challenging to see such quantum
effects in the interaction of the condensate with a series of
defects created with light.

In conclusion, we have experimentally studied the 1D expansion of
a BEC in a linear optical guide in the presence of a random
potential. The random potential is created by a speckle field that
has been switched on either adiabatically or abruptly after some
initial expansion of the condensate. In both the cases we have
observed a halted expansion of the atomic cloud and the
suppression of the center of mass motion. In order to understand
the role played by randomness in this behaviour we have repeated
the same experiments with a single defect created by a tightly
focused laser beam. The experimental results are confirmed by GPE
simulations showing that the suppressed expansion is mainly caused
by trapping in the deepest wells of the potentials. We have also
carried out a detailed investigation of a BEC performing center of
mass oscillations in a harmonic trap interacting with a single
optical well. The reported measurements indicate that the
diffraction limited size of defects created by light poses severe
restrictions on the range of parameters necessary to observe
quantum reflection/transmission that is at the basis of Anderson
localization of matterwaves.

This work has been supported by the EU Contracts No.
HPRN-CT-2000-00125, INFM PRA ``Photon Matter'' and MIUR FIRB 2001.
We thank all the LENS Quantum Gases group for useful discussions.


\begin{thebibliography}{99}
\bibitem{autocorrelazione} The spatial autocorrelation function of the
speckles potential is a Gaussian with $\sigma_z=5.7 \pm 0.4
$~$\mu$m.
\bibitem{teoria} A detailed discussion will be presented in a future publication.
\bibitem{nota2} We performed similar experiment letting the
condensate ``collide'' with the single well during expansion in
the optical linear guide, but in this case the elongated atomic
cloud could mask the eventual reflection from the defect.
\end{thebibliography}
\end{document}